\documentclass[onecolumn,amsmath,amssymb,superscriptaddress,nofootinbib,11pt,a4paper]{revtex4}

\pdfoutput=1
\usepackage[T1]{fontenc}
\usepackage{CJK}
\usepackage{xcolor}
\usepackage{amsfonts}
\usepackage{epstopdf}
\usepackage{wrapfig} 
\usepackage{subcaption}
\usepackage{graphicx}  
\usepackage{dcolumn}   
\usepackage{bm}
\usepackage{float}
\usepackage[T1]{fontenc}
\usepackage{CJK}
\usepackage{xcolor}
\usepackage{amsfonts}
\usepackage{epstopdf}
\usepackage{wrapfig} 
\usepackage{subcaption} 
\usepackage{graphicx}  
\usepackage{dcolumn}   
\usepackage{bm}
\usepackage{float}
\usepackage[utf8]{inputenc}
\usepackage[T1]{fontenc}

\begin{document}

\title{Hotspot Images Driven by Magnetic Reconnection in Kerr-Sen black hole}

\author{Ke Wang}
\email{kkwwang2025@163.com}
\affiliation{School of Material Science and Engineering, Chongqing Jiaotong University, \\Chongqing 400074, China}

\author{Xiao-Xiong Zeng\footnote{Electronic address: xxzengphysics@163.com.  (Corresponding author)}}
\affiliation{College of Physics and Electronic Engineering, Chongqing Normal University, \\Chongqing 401331, China}

\begin{abstract}
{In the Kerr-Sen black hole, this study investigates the changes in hotspot images before and after the occurrence of magnetic reconnection. After reviewing the Comisso-Asenjo magnetic reconnection process and introducing the hotspot imaging method, we examine the temporal evolution of hotspot intensity, including  when energy extraction occurs, when it does not occur, and when the observer's azimuthal angle is altered. We also discuss the influence of the black hole's expansion parameter and spin on hotspot imaging. The results indicate that the first flare may serve as a potential signature of ongoing energy extraction; changing the observer's azimuthal angle may alter the time interval between the first and second flares; a larger expansion parameter makes it more difficult to identify the energy extraction signal, and a higher spin also makes it more challenging to detect the energy extraction signal.
}
\end{abstract}

\maketitle
 \newpage
\section{Introduction}
Observational evidence indicates the presence of a supermassive black hole at the center of the Milky Way \cite{13,14}, and near-infrared flare events have been observed near the event horizon of the Sgr A* black hole \cite{15}. These flares can be detected across all wavelengths. There are several methods to study the source of near-infrared flares, and one common approach involves modeling hot spots orbiting the central black hole \cite{16,17,18,19,20,11}. This highlights the importance of hot spot imaging, as the flares it produces hold significant value for astronomical observations.

Recently, research on magnetic reconnection processes within the framework of general relativity has advanced rapidly \cite{33,34,35,36}. Since magnetic reconnection events can rapidly release energy stored in magnetic fields and convert it into kinetic and internal energy of magnetized fluids, this mechanism can serve as a candidate model for hot spots, potentially detectable by our telescopes. Furthermore, because magnetic reconnection can accelerate plasma to near-light speeds, when such an event occurs within the black hole's ergosphere, it can act as an ignition mechanism for the Penrose process \cite{12}, thereby enabling the extraction of the black hole's rotational energy. Observational evidence for the Penrose process remains scarce, as the process occurs within the ergosphere, very close to the event horizon, making it challenging for current telescope technology to resolve. Theoretically, studies on how to test the Penrose process via hot spots are also highly limited \cite{48}. Current observations suggest that the flares from the Sgr A* black hole originate from hot spots orbiting at approximately 9 gravitational radii \cite{20}. Some studies indicate that these flares may correspond to hot spots in orbit within active galactic nuclei (AGN) \cite{45,46,47}. Therefore, with advancing technology, astronomical telescopes are expected to observe the ergosphere region and detect signals of energy extraction. Thus, we simulate the particles involved in this process as hot spots, observe them to identify flare phenomena, and determine potential energy extraction events, thereby assessing the impact on hot spot imaging. This magnetic reconnection mechanism for extracting black hole energy was established by Asenjo and Comisso \cite{1}. This energy extraction mechanism has been extensively generalized to other rotating black holes \cite{21,22,23,24,40,25}. For our hot spot imaging, the magnetic reconnection mechanism employed is precisely the Comisso-Asenjo process.

The Kerr-Sen solution originates from low-energy effective string theory \cite{26,27} and encompasses not only the gravitational field but also two additional fields in the form of a dilaton field and a U(1) gauge field similar to the electromagnetic field. It closely resembles the Kerr-Newman solution, as both possess charge, but they are fundamentally distinct. The charge of the Kerr-Newman black hole comes from the classical electromagnetic field, representing charge in Maxwell's theory. In contrast, the charge of the Kerr-Sen black hole stems from the $U(1)$ gauge field in string theory and is always accompanied by a dilaton field. The Kerr-Sen metric is not a solution to the Einstein-Maxwell equations. Many properties of the Kerr-Sen black hole have been extensively studied \cite{28,29,30,31,32}. This paper employs the hot spot imaging method to observe the motion of plasma after (and briefly before) the occurrence of the Comisso-Asenjo process in a Kerr-Sen black hole. Our research indicates that during the hot spot imaging process, observers would detect three flares within the observation period. The first flare may serve as a potential signature of ongoing energy extraction; changing the observer's azimuthal angle may alter the time interval between the first and second flares; a larger expansion parameter results in a lower total intensity of the first flare, thereby making it more difficult to identify the energy extraction signal; a higher spin also leads to a lower total intensity of the first flare and a shorter time interval between the first and second flares, further complicating the identification of the energy extraction signal.

The remainder of this paper is organized as follows. In Section 2, we provide a brief description of the magnetic reconnection process in the Kerr-Sen black hole. Section 3 offers a concise introduction to the hot spot model and the imaging method. In Section 4, we present the observational results, including cases with energy extraction, without energy extraction, and with varying observer azimuthal angles. Section 5 analyzes the influence of the expansion parameter and spin on the hot spot intensity. We conclude the paper in Section 6.

\section{Comisso-Asenjo process in the Kerr-Sen Black Hole}

In this section, we begin by reviewing the magnetic reconnection process described in reference \cite{1}, namely the Comisso-Asenjo process. We use natural units $(c=G=1)$. The line element of the Kerr-Sen metric in Boyer-Lindquist (BL) coordinates is given by \cite{32,2,3}
\begin{equation}
ds^{2} = -(1-\frac{2Mr}{\Sigma})dt^{2}+\frac{\Sigma}{\Delta}dr^{2}+\Sigma d \theta^{2}-\frac{4aMr}{\Sigma}\sin^{2}\theta dtd\phi+\sin^{2}\theta\left(r(r+r_{0})+a^{2}+\frac{2Mra^{2}\sin^{2}\theta }{\Sigma}\right)d\phi^{2},    
\end{equation}
where
\begin{equation}
\Sigma=r(r+r_{0})+a^{2}\cos^{2}\theta,\Delta=r(r+r_{0})+a^{2}-2Mr.    
\end{equation}
Here, $M$ represents the black hole mass, $a$ is the spin, $r_0 = Q^2/M$ is the expansion parameter, and $Q$ denotes the black hole charge. The event horizon of the black hole is determined by $\Delta = 0$, with solutions given by
\begin{equation}
r_h = M - \frac{r_0}{2} \pm \sqrt{\left( M - \frac{r_0}{2} \right)^2 - a^2}.  \label{3} 
\end{equation}
Here, the $+$ and $-$ signs correspond to the event horizon and the inner horizon, respectively. The boundary of the ergosphere is determined by $g_{tt} = 0$. For a unit mass $M = 1$, the solution is
\begin{equation}
r_E = \frac{2 - r_0 \pm \sqrt{(r_0 - 2)^2 - 4 a^2 \cos^2 \theta}}{2} .\label{4}   
\end{equation}
Here, the $+$ and $-$ signs represent the outer ergosphere and inner ergosphere, respectively. On the other hand, the intrinsic frame-dragging effect of rapidly rotating black holes induces an antiparallel magnetic field line configuration near the equatorial plane \cite{4,5,6}. The sudden reversal of magnetic field lines dynamically generates an equatorial current sheet. Once the aspect ratio of this current sheet surpasses a critical value, plasma instabilities cause it to rupture \cite{7,8}. Following this, the formation of plasma flux ropes drives fast magnetic reconnection, rapidly converting stored magnetic energy into kinetic energy of plasma particles. In particular, the equatorial current sheet is prone to the plasmoid instability, resulting in its fragmentation and the creation of multiple X-points.  Among these X-points, the dominant X-point is defined as the one located at the intersection of the separatrix that encloses the global reconnection outflow, and it governs the overall reconnection dynamics. As magnetic reconnection proceeds, plasma is expelled from the reconnection layer, and the release of magnetic tension drives plasma outflow. Simultaneously, the frame-dragging of spacetime continually stretches the magnetic field lines, repeatedly forming current sheets and perpetuating a cycle of magnetic reconnection. Within this process, a fraction of the plasma is accelerated outward, while the remainder is decelerated. When occurring within the ergosphere, the decelerated plasma can be captured in negative-energy orbits and ultimately captured by the black hole. In accordance with the Penrose process, this leads to the ejection of the accelerated plasma to infinity, resulting in net energy extraction from the black hole. This paper employs the hot spot imaging method to observe the motion of the plasma during this process. In this study, we adopt a unit mass $M=1$.

Based on the simplifying assumptions in reference \cite{1}, the current sheet moves in a Keplerian circular orbit within the equatorial plane. The Keplerian angular velocity is given by
\begin{equation}
\Omega = -\frac{\partial_{r} g_{t \phi}}{\partial_{r} g_{\phi \phi}} \pm \frac{\sqrt{\left(\partial_{r} g_{t \phi}\right)^{2} - \left(\partial_{r} g_{t t}\right)\left(\partial_{r} g_{\phi \phi}\right)}}{\partial_{r} g_{\phi \phi}}.   
\end{equation}
Here, the $+$ and $-$ signs represent the prograde and retrograde orbits, respectively. This orbit can extend from infinity all the way to the photon sphere. If the current sheet lies within the ergosphere, only the prograde orbit exists. Because the retrograde photon sphere is located outside the ergosphere. The line element of the spacetime can be expressed in the $3+1$ formalism as
\begin{equation}
ds^{2}=g_{\mu\nu}dx^{\mu}dx^{\nu}=-\alpha^{2}dt^{2}+\sum_{i=1}^{3}\bigl {(}\sqrt{g_{ii}}dx^{i}-\alpha\beta^{i}dt\bigr{)}^{2},   
\end{equation}
where
\begin{equation}
\alpha=\sqrt{-g_{tt}+\frac{g_{t\phi}^{2}}{g_{\phi\phi}}}\,,\beta^{i}=\delta_{i\phi}\frac{\sqrt{g_{\phi\phi}}\,\omega^{\phi}} {\alpha},\omega^{\phi}=-g_{t\phi}/g_{\phi\phi}  .  
\end{equation}
A tetrad in the Zero Angular Momentum Observer (ZAMO) reference frame is established as
\begin{equation}
\hat{e}_0 = \frac{1}{\alpha}(\partial_0 + \omega^\phi \partial_3), \hat{e}_i = \frac{1}{\sqrt{g_{ii}}} \partial_i  . 
\end{equation}
Therefore, the transformation relation for the four-velocity between the BL reference frame and the ZAMO reference frame is given by
\begin{equation}
\hat{U}^{\mu}=\hat{\gamma}\left\{1, \hat{v}^{(r)}, 0, \hat{v}^{(\phi)}\right\}=\left\{\frac{E-\omega^{\phi} L}{\alpha}, \sqrt{g_{r r}} U^{r}, 0, \frac{L}{\sqrt{g_{\phi \phi}}}\right\} ,\label{9}
\end{equation}
where
\begin{equation}
\hat{v}=\sqrt{\left(\hat{v}^{(r)}\right)^{2}+\left(\hat{v}^{(\phi)}\right)^{2}},\hat{\gamma}=(1-\left.\hat{v}^{2}\right)^{-1 / 2}.
\end{equation}
The Keplerian velocity in the ZAMO reference frame is
\begin{equation}
\hat{v}_{K}=\frac{1}{\alpha}\left(\sqrt{g_{\phi \phi}} \Omega-\alpha \beta^{\phi}\right)       .  
\end{equation}
The four-velocity of the current sheet is then
\begin{equation}
\hat{u}_{K} = \hat{\gamma}_{K}(1,0,0,\hat{v}_{K}).    
\end{equation}
According to equation \eqref{9}, the energy and angular momentum of the current sheet in the BL coordinate system are
\begin{equation}
L=\sqrt{g_{\phi \phi}}\hat{\gamma}_{K}\hat{v}_{K},E=\alpha\hat{\gamma}_{K}+\omega^\phi L.\label{13}
\end{equation}
To satisfy the circular orbit condition, the values of 
$E$ and $L$ in equation \eqref{13} must be real. A tetrad in the fluid rest frame is established as
\begin{equation}
\sigma_{0} = \hat{\gamma}_{K}(\hat{e}_{0} + \hat{v}_{K}\hat{e}_{3}),
\sigma_{1} = \hat{e}_{1},
\sigma_{2} = \hat{e}_{2},
\sigma_{3} = \hat{\gamma}_{K}(\hat{v}_{K}\hat{e}_{0} + \hat{e}_{3}). 
\end{equation}
Assuming the azimuthal angle of the magnetic field lines in the fluid rest frame is $\xi$, the four-velocity of the plasma ejected from the reconnection layer in the fluid rest frame is
\begin{equation}
u^{\prime\mu} = \gamma_{out}[\sigma_{0}^{\mu} \pm v_{out}(\cos\xi\sigma_{3}^{\mu} + \sin\xi\sigma_{1}^{\mu})],   
\end{equation}
Here, the $+$ and $-$ signs correspond to the accelerated and decelerated plasmas, respectively. $v_{out}$ denotes the magnitude of the outflow velocity observed in the fluid rest frame, which satisfies \cite{1}
\begin{equation}
v_{ out }=\sqrt{\frac{\sigma}{1+\sigma}}  ,  
\end{equation}
where $\sigma$ is the upstream magnetization parameter. In the ZAMO reference frame, the four-velocity of the plasma ejected from the reconnection layer is
\begin{equation}
\hat{u}^{\mu} = \gamma_{out}[\hat{\gamma}_{K}(\hat{e}_{0}^{\mu} + \hat{v}_{K}\hat{e}_{3}^{\mu}) \pm v_{out}(\cos\xi(\hat{\gamma}_{K}(\hat{v}_{K}\hat{e}_{0}^{\mu} + \hat{e}_{3}^{\mu})) + \sin\xi\hat{e}_{1}^{\mu})] = \hat{\gamma}_{out}(1,\hat{v}^{r},0,\hat{v}^{\phi}),    
\end{equation}
where
\begin{equation}
\hat{\gamma}_{out} = \gamma_{out}\hat{\gamma}_{K}(1 \pm \hat{v}_{K}v_{out}\cos\xi), 
\hat{v}^{r} = \pm \frac{v_{out}\sin\xi}{\hat{\gamma}_{K}(1 \pm \hat{v}_{K}v_{out}\cos\xi)},
\hat{v}^{\phi} = \frac{\hat{v}_{K} \pm v_{out}\cos\xi}{1 \pm \hat{v}_{K}v_{out}\cos\xi} .   
\end{equation}
According to equation \eqref{9}, the energy, angular momentum, and radial component of the four-momentum of the plasma ejected from the reconnection layer in the BL coordinate system are
\begin{equation}
\mathcal{L}=\sqrt{g_{\phi \phi}}\hat{\gamma}_{out}\hat{v}^{\phi}=p_\phi,\mathcal{E}=\alpha\hat{\gamma}_{out}+\omega^\phi \mathcal{L}=-p_t, p_r= \hat{\gamma}_{out}\hat{v}^{r}  \sqrt{g_{rr}}.\label{19}
\end{equation}
Equation \eqref{19} provides the equations of motion for the ejected plasma. This expression is crucial for hot spot imaging, used to observe the motion of accelerated or decelerated plasmas. Assuming the magnetic reconnection process is highly efficient, electromagnetic field energy can be neglected, and the plasma is treated as an adiabatic incompressible fluid. According to \cite{1}, the energy-at-infinity per unit enthalpy of the plasma ejected from the reconnection layer can be approximated as
\begin{equation}
\varepsilon_{\pm}=\alpha \hat{\gamma}_{K}\left[\left(1+\beta^{\phi} \hat{v}_{K}\right)\sqrt{1+\sigma} \pm \cos \xi\left(\beta^{\phi}+\hat{v}_{K}\right) \sqrt{\sigma} - \frac{1}{4} \frac{\sqrt{1+\sigma} \mp \cos{\xi} \hat{v}_{K} \sqrt{\sigma}}{\hat{\gamma}_{K}^{2}\left(1+\sigma-\cos ^{2}{\xi} \hat{v}_{K}^{2} \sigma\right)}\right].\label{20} 
\end{equation}
For energy extraction to occur, the following two criteria must be satisfied \cite{1}
\begin{align}
\varepsilon_{-}<0, \varepsilon_{+}>0.
\end{align}
where we assume the plasma is relativistically hot and adopt a polytropic index of 4/3.
\section{Hot Spot Model and Imaging Method}

Here, we outline the hot spot model and the relevant imaging framework. As in Ref. \cite{48}, We forgo an investigation into how radiation mechanisms influence imaging or flares, as our approach does not account for the specifics of the emission spectrum. Instead, the model assumes the hot spot emits isotropically with a frequency-independent, flat spectrum, effectively simulating a broadband source. The plasmoid is thus represented as a transparent hot spot whose emission rate is described by a normal distribution
\begin{equation}
J_{\nu_s} = \exp\left( \frac{ -x^2 }{ 2 \cdot s^2 } \right).
\end{equation}
Here, $J_{\nu_{s}}$ is the emissivity, $s$ is the standard deviation of the normal distribution, and $x$ is the distance from the center of the hot spot. Imaging a moving hot spot requires tracing the path of the light source and simulating its radiative transfer. As specified by equation \eqref{19}, the trajectory of the light source is determined by numerically integrating its geodesic equations. For the radiative transfer component, we utilize a backward ray-tracing technique coupled with a fisheye camera model. The specific computational approach is detailed in Appendix B of reference \cite{9}. The camera is placed in the ZAMO reference frame, and its tetrad is given by 
\begin{equation}
\hat{e}_{(0)}=\frac{g_{\phi\phi}\,\partial_{0}-g_{\phi t}\,\partial_{3}}{\sqrt{g_{\phi\phi}\Big(g_{\phi t}^{2}-g_{\phi\phi}g_{tt}\Big)}},\hat{e }_{(1)}=-\frac{\partial_{1}}{\sqrt{g_{rr}}}, \hat{e}_{(2)}=\frac{\partial_{2}}{\sqrt{g_{\theta\theta}}}\,, \hat{e}_{(3)}=-\frac{\partial_{3}}{\sqrt{g_{\phi\phi}}}.   
\end{equation}
The intensity on the image plane is determined by the radiative transfer equation \cite{10,Zeng:2025kqw,Zeng:2021dlj,Zeng:2021mok}
\begin{equation}
\frac{d}{d\lambda}\left(\frac{I_{\nu}}{\nu^{3}}\right)=\frac{J_{\nu}-k_{\nu}I_{\nu}}{\nu^{2}}\,,    
\end{equation}
where $\lambda$ is the affine parameter along the null geodesic, and $I_{\nu},J_{\nu}$ and $k_{\nu}$ represent the specific intensity, emissivity, and absorption coefficient at frequency $\nu$, respectively. In this study, the absorption coefficient is neglected. Our numerical scheme produces a series of snapshots where the recorded photon arrival time corresponds to the sum of the source's orbital time and the light travel time to the observer. The projected position of the flux centroid on the camera plane is also determined for each individual frame. In accordance with the camera framework adopted from \cite{9}, the flux projected onto a specific $(i,j)$-th pixel is provided by \cite{11}
\begin{equation}
F(i,j)=I_{\nu}(i,j)S_{0}\cos\left[2\arctan\left(\frac{1}{n}\tan\left(\frac{\alpha_{ fov}}{2}\right)\sqrt{\left(i-\frac{n+1}{2}\right)^{2}+\left(j-\frac{n+1}{2}\right) ^{2}}\right)\right].    
\end{equation}
Here, $S_0$ is the area of a single pixel, $n$ is the total number of pixels along the horizontal or vertical axis, and $i,j$ range from 1 to $n$. $\alpha_{ fov}$ is the camera's field of view. After obtaining this data, the centroid position $\vec{x}_{c}(t)$ of each image can be calculated using the following formula
\begin{equation}
\vec{x}_{c}(t)=\frac{\sum_{i,j}\vec{x}(i,j)F(i,j)}{\sum_{i,j}F(i,j)}\,,    
\end{equation}
where $\vec{x}(i,j)$ represents the spatial coordinates of the $(i,j)$-th pixel. The total flux in a given snapshot, denoted as $\sum_{i,j}F(i,j)$, is referred to as the flux associated with $\vec{x}_{c}(t)$ By employing our model and the computational framework described above, we can ultimately track the evolution of the brightness centroid position and the corresponding flux over time.
\section{Observational Results}
\subsection{Observations with Energy Extraction}
First, we set the parameters as $\sigma=30$, $\xi=\pi/20$, $a=0.9$, $s=0.2$, with the observer's azimuthal angle $\phi_0=\pi/2$, inclination angle $\theta_0=\pi/10$, and radial distance set to 200. The radial distance of the dominant X-point is $r_X=1.6$, the expansion parameter related to charge is $r_0=0.1$, and the initial time when the hot spot first appears on the screen is $t=0$. According to equations \eqref{3} and \eqref{4}, the current sheet is located within the ergosphere ($r_{h+}=1.254 < r_X < r_{E+}=1.9$). Since only prograde orbits exist within the ergosphere, we use the prograde orbit for calculations. This yields $E=0.847052$ and $L=2.00447$, satisfying the condition for real values under circular orbits. Based on \eqref{20}, we obtain $\varepsilon_{-}=-0.4254$ and $\varepsilon_{+}=9.8258$, which satisfy the energy extraction condition. We present the observational results as follows.
\begin{figure}[!h]
  \centering
  \begin{subfigure}{0.22\textwidth}
    \centering
    \includegraphics[width=\linewidth]{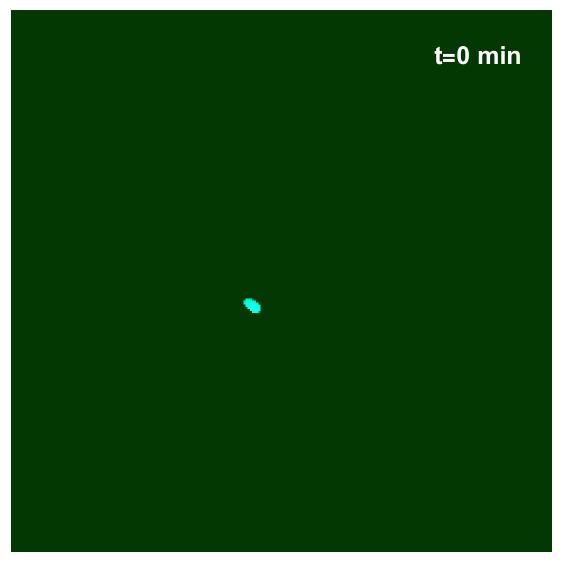} 
    \label{fig:1a}
  \end{subfigure}
  \begin{subfigure}{0.22\textwidth}
    \centering
    \includegraphics[width=\linewidth]{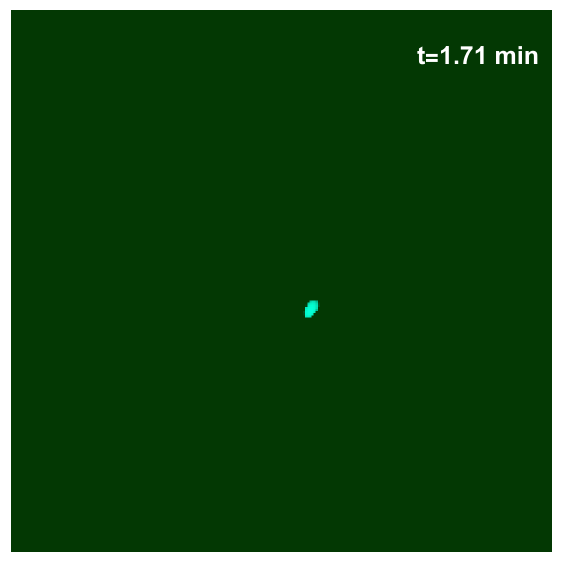} 
    \label{fig:1b}
  \end{subfigure}
 \begin{subfigure}{0.22\textwidth}
    \centering
    \includegraphics[width=\linewidth]{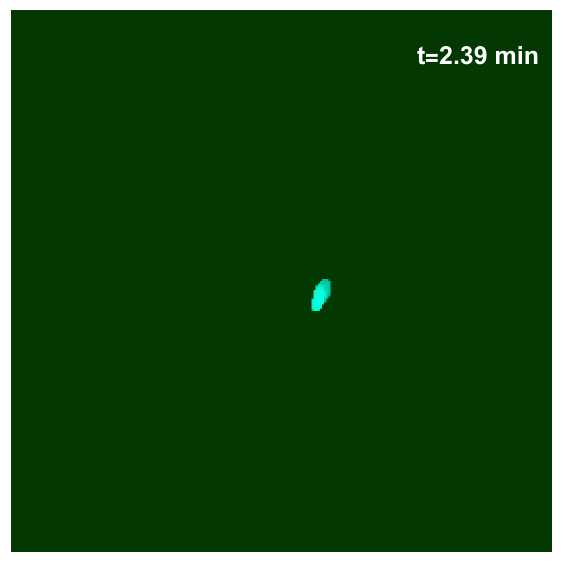}  
    \label{fig:1c}
  \end{subfigure}
 \begin{subfigure}{0.22\textwidth}
    \centering
    \includegraphics[width=\linewidth]{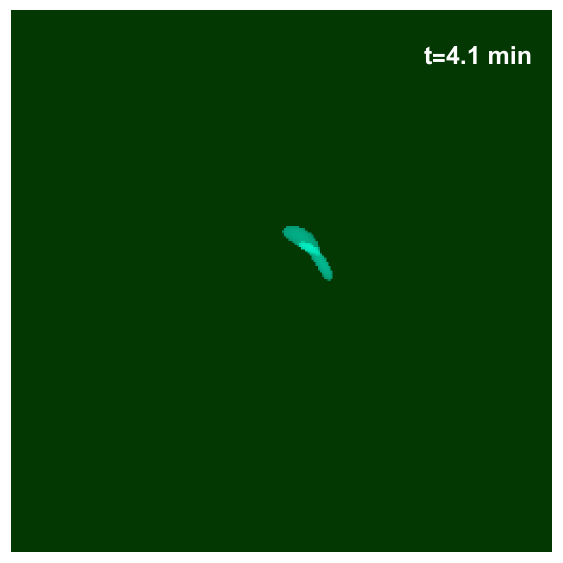} 
    \label{fig:1d}
  \end{subfigure}
 \begin{subfigure}{0.22\textwidth}
    \centering
    \includegraphics[width=\linewidth]{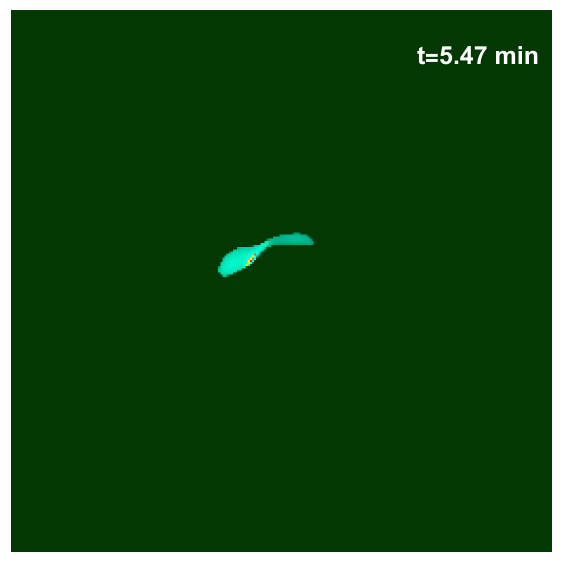}  
    \label{fig:1e}
  \end{subfigure}
 \begin{subfigure}{0.22\textwidth}
    \centering
    \includegraphics[width=\linewidth]{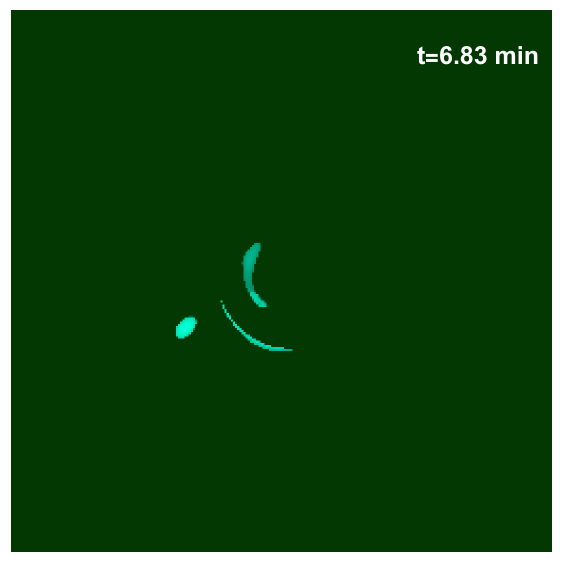} 
    \label{fig:1f}
  \end{subfigure} 
 \begin{subfigure}{0.22\textwidth}
    \centering
    \includegraphics[width=\linewidth]{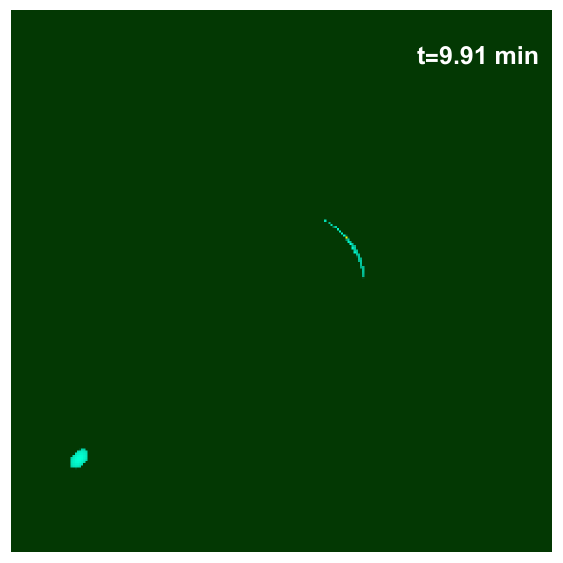} 
    \label{fig:1g}
  \end{subfigure} 
  \begin{subfigure}{0.22\textwidth}
    \centering
    \includegraphics[width=\linewidth]{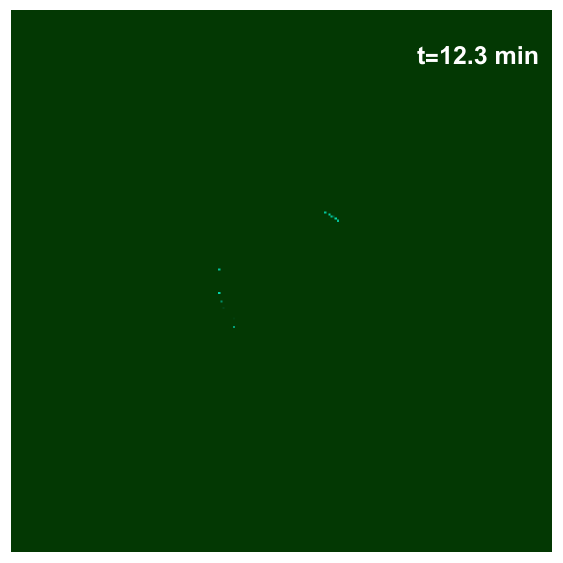} 
    \label{fig:1h}
  \end{subfigure} 
  \caption{Time evolution of the plasma hot spot intensity distribution in the Kerr-Sen black hole}
  \label{fig:1}
\end{figure}

According to Fig. \ref{fig:1}, at $t=0$, a hot spot in Keplerian motion appears on the screen. At $t=1.71$ min, the transient magnetic reconnection process is just completed. By $t=2.39$ min, the hot spots of the accelerated and decelerated plasmas become distinguishable on the screen. At $t=4.1$ min, the hot spot is located at the first flare. By $t=5.47$ min, the hot spot reaches the second flare. At $t=6.83$ min, it is clearly visible that the accelerated plasma moves toward infinity, while the decelerated plasma hovers near the event horizon. Secondary or higher-order images also appear on the screen. By $t=9.91$ min, the decelerated plasma has disappeared from the screen, meaning it has fallen into the event horizon, and the hot spot is now at the third flare. At $t=12.3$ min, both types of plasma have completely left the screen, leaving only secondary or higher-order images.
\begin{figure}[!h]
  \centering
  \begin{subfigure}{0.32\textwidth}
    \centering
    \includegraphics[width=\linewidth]{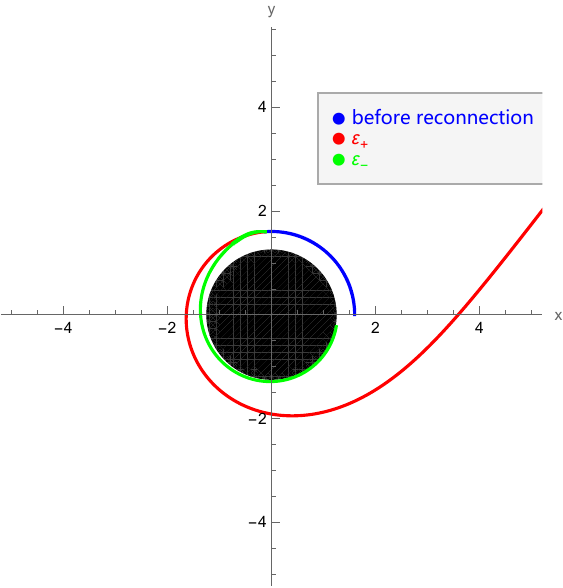}
    \label{fig:2a}
  \end{subfigure}
  \begin{subfigure}{0.32\textwidth}
    \centering
    \includegraphics[width=\linewidth]{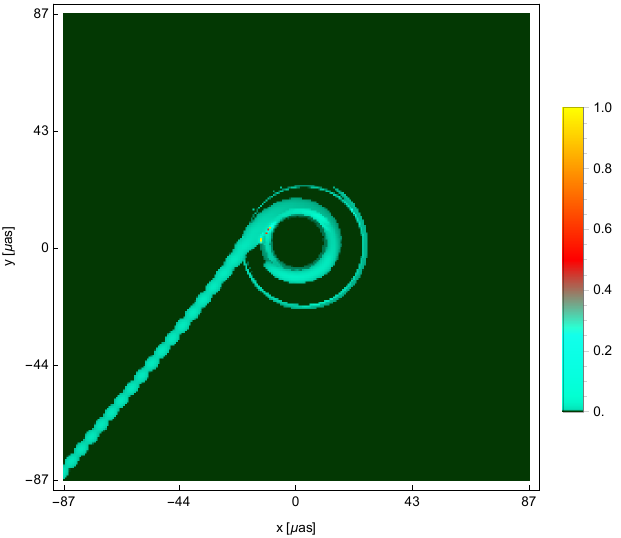} 
    \label{fig:2b}  
  \end{subfigure}
  \begin{subfigure}{0.32\textwidth}
    \centering
    \includegraphics[width=\linewidth]{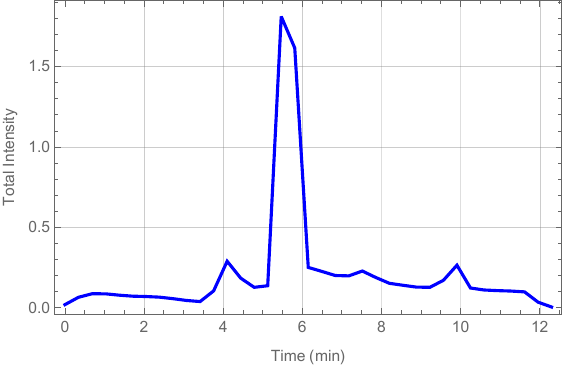} 
    \label{fig:2c} 
  \end{subfigure}
  \caption{The left panel displays the trajectory of the plasmoid in a two-dimensional Cartesian coordinate system. The blue curve represents the case without magnetic reconnection, the red curve represents the accelerated plasma, and the green curve represents the decelerated plasma. The black circle denotes the black hole. The middle panel depicts the observed hot spot intensity distribution, showing the time-averaged radiation intensity on the observation plane, normalized by $I/I_{MAX}$. The outer ring in the middle panel represents secondary or higher-order images. The inner ring and the tailed stripe in the middle panel are the primary images. The right panel shows the light curve of the hot spot emission, revealing the variation of the total flux with observation time.}
  \label{fig:2}
\end{figure}

\begin{figure}[!h]
  \centering
  \begin{subfigure}{0.32\textwidth}
    \centering
    \includegraphics[width=\linewidth]{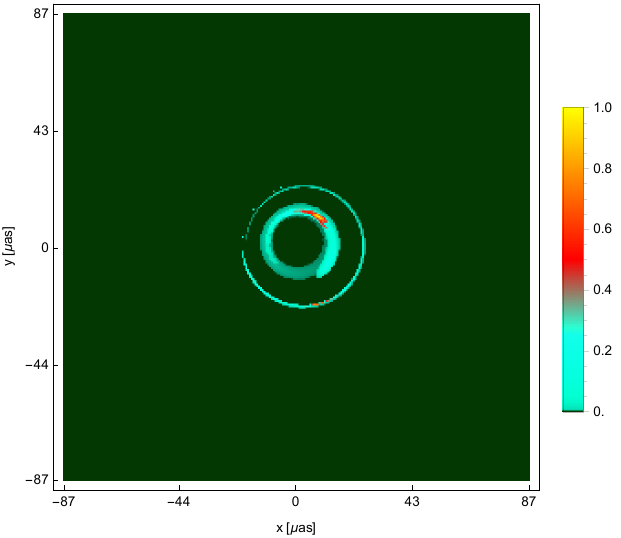}
    \label{fig:3a}
  \end{subfigure}
  \begin{subfigure}{0.32\textwidth}
    \centering
    \includegraphics[width=\linewidth]{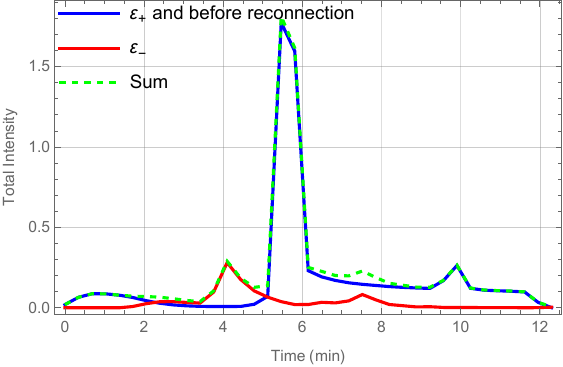} 
    \label{fig:3b}  
  \end{subfigure}
  \begin{subfigure}{0.17\textwidth}
    \centering
    \includegraphics[width=\linewidth]{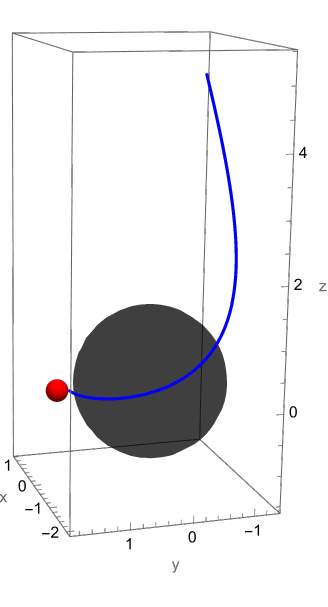} 
    \label{fig:3c}  
  \end{subfigure}
  \caption{The left panel illustrates the normalized hot spot intensity distribution generated solely by the decelerated plasma. The middle panel shows light curves, the blue solid line represents $\varepsilon_{+}$ and the before reconnection intensity, the red solid line represents the intensity from only $\varepsilon_{-}$, and the green dashed line represents the total light curve observed. The right panel displays the photon trajectories corresponding to the first flare.}
  \label{fig:3}
\end{figure}

According to the light curve in the right panel of Fig. \ref{fig:2}, it can be observed that three flares are detected within the observation period: the first and third are faint, while the second is bright.\footnote{At the 7.5-minute mark, the light curve shows a slight bump, which, as seen in the middle panel of Fig. \ref{fig:3}, originates from $\varepsilon_-$. This bump is very faint, and we do not classify it as a flare. Even if it were considered a flare, it would not affect the analysis of the first flare in the light curve.} To determine the origins of these three flares, we plotted light curves for three cases: the hot spot image generated solely by the decelerated plasma ($\varepsilon_-$), the total intensity from $\varepsilon_+$ and the before reconnection intensity, and the total intensity observed by the viewer. These are shown in Fig. \ref{fig:3}. It can be seen that the first flare originates from the negative-energy plasmoid, while the subsequent two flares come from the positive-energy plasmoid. Specifically, the flare from the negative-energy plasmoid arises from the primary image, as clearly shown in the right panel of Fig. \ref{fig:3}. The brightest flare occurs because the accelerated plasmoid gains substantial kinetic energy and experiences significant Doppler blueshift. The first flare originates from the Doppler blueshift of the primary image of the decelerated plasmoid. The third flare results from the Doppler blueshift effect on the secondary image of the accelerated plasmoid. For the detailed mechanism of flare production, see Appendix A of Ref. \cite{48}. In Ref. \cite{48}, they also intentionally introduced a control case with plasma of $\varepsilon_->0$ but having the same magnitude of energy. This plasma did not produce a flare, demonstrating an asymmetry between the $\varepsilon_-<0$ plasma that falls into the black hole and the $\varepsilon_->0$ plasma. These results indicate that the first flare has the potential to serve as a potential signature of ongoing energy extraction via the Penrose process.

\subsection{Observational Results without Energy Extraction}
\begin{figure}[!h]
  \centering
  \begin{subfigure}{0.35\textwidth}
    \centering
    \includegraphics[width=\linewidth]{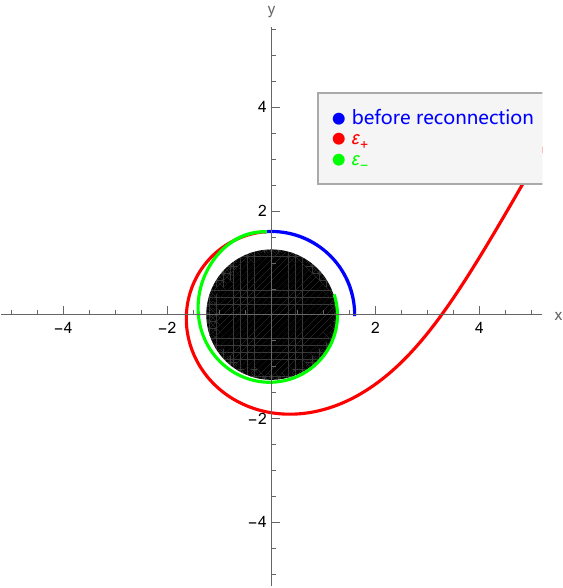}
    \label{fig:4a}
   \end{subfigure}
   \begin{subfigure}{0.35\textwidth}
    \centering
    \includegraphics[width=\linewidth]{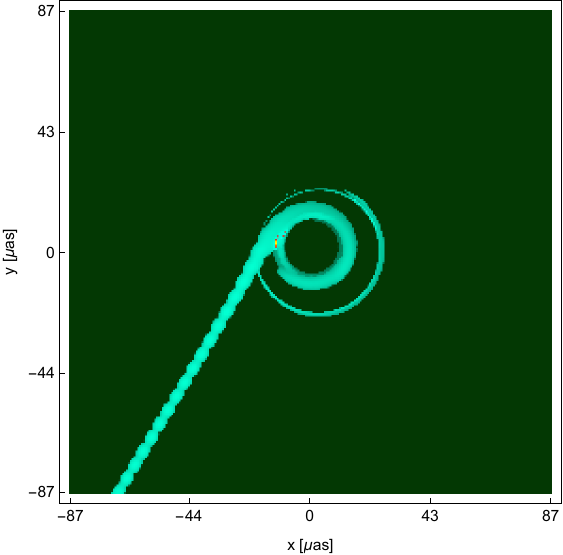}
    \label{fig:4b}
   \end{subfigure}
   \begin{subfigure}{0.35\textwidth}
    \centering
    \includegraphics[width=\linewidth]{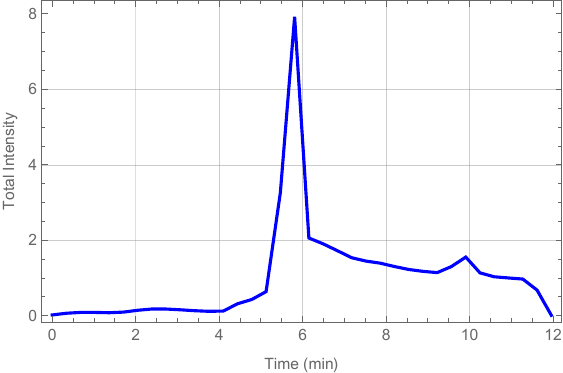}
    \label{fig:4c}
   \end{subfigure}
  \begin{subfigure}{0.35\textwidth}
    \centering
    \includegraphics[width=\linewidth]{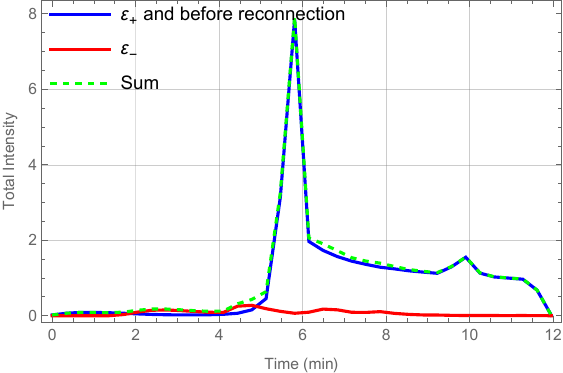}
    \label{fig:4d}
   \end{subfigure}
\caption{In the absence of energy extraction, the top-left panel shows the trajectory of the plasmoid, the top-right panel depicts the normalized intensity distribution of the hot spot observed, the bottom-left panel displays the light curve of the hot spot emission, and the bottom-right panel presents the three types of light curves.}
  \label{fig:4}
\end{figure}

We conducted a set of comparative observations. Here, we only changed the magnetization parameter $\sigma$ to 3, while keeping all other conditions identical to the previous subsection. In this case, $\varepsilon_{-}=0.01683$ and $\varepsilon_{+}=3.2961$, which do not satisfy the energy extraction condition. The observational results are shown in Fig. \ref{fig:4}.

From the top-left panel of Fig. \ref{fig:4} and the left panel of Fig. \ref{fig:2}, it can be observed that the trajectories are identical when magnetic reconnection does not occur. This is because changing the magnetization parameter does not affect the motion of the Keplerian orbit. From the bottom-right panel of Fig. \ref{fig:4}, it can be seen that without energy extraction, only two flares from $\varepsilon_{+}$ are perceptible, while the flare (or slight bump) from $\varepsilon_{-}$ is almost imperceptible. Although this could be attributed to the change in the magnetization parameter, it also indicates that the flare produced by $\varepsilon_{-}$ is unique and may potentially serve as a signature of ongoing energy extraction via the Penrose process.
\subsection{Observational Results with Varying Observer Azimuthal Angle}
\begin{figure}[!h]
  \centering
  \begin{subfigure}{0.32\textwidth}
    \centering
    \includegraphics[width=\linewidth]{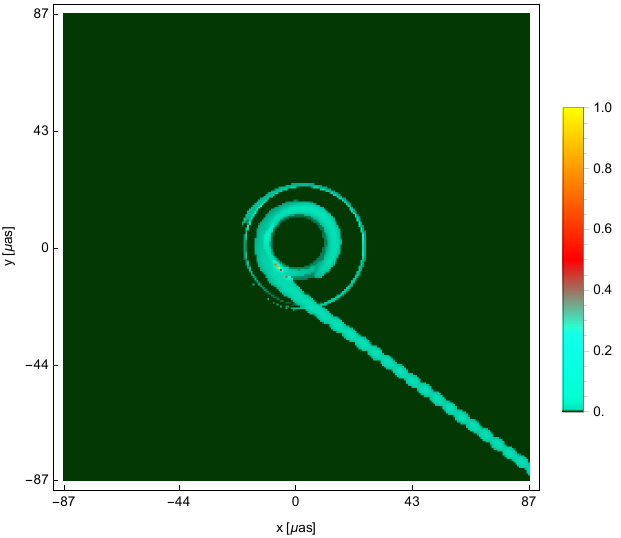}
    \label{fig:7a}
  \end{subfigure}
  \begin{subfigure}{0.32\textwidth}
    \centering
    \includegraphics[width=\linewidth]{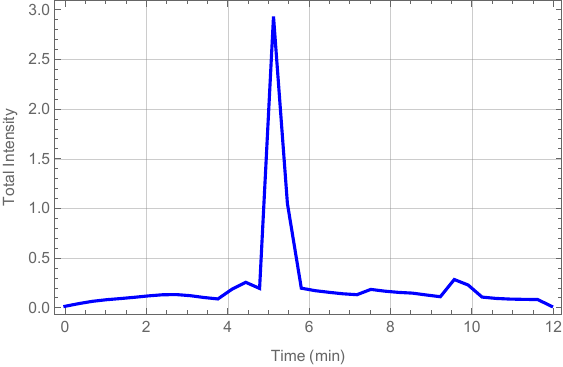} 
    \label{fig:7b}  
  \end{subfigure}
  \begin{subfigure}{0.32\textwidth}
    \centering
    \includegraphics[width=\linewidth]{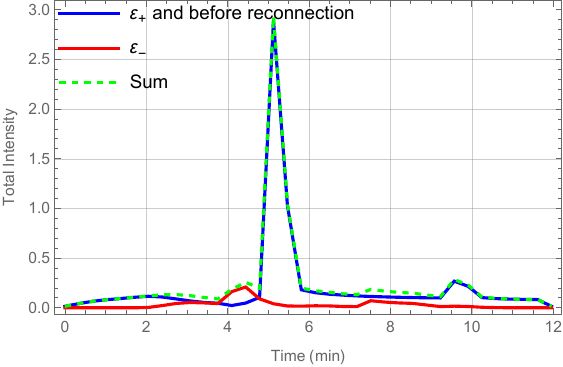} 
    \label{fig:7c}  
  \end{subfigure}
  \caption{When $\phi_0=0$, the left panel depicts the normalized intensity distribution of the hot spot observed, the middle panel shows the light curve of the hot spot emission, and the right panel displays the light curves for the three cases.}
  \label{fig:7}
\end{figure}

Next, we set the observer's azimuthal angle to $\phi_0=0$, while keeping all other parameters the same as in Section 4.A. The observational results are presented in Fig. \ref{fig:7}.
Comparing the middle panel of Fig. \ref{fig:3} and the right panel of Fig. \ref{fig:7}, it can be observed that after changing the observer's azimuthal angle, some changes occur in the light curves. The $\varepsilon_{-}$ still produces the first flare, and the total intensity of the first flare in both cases shows little difference, approximately 0.25. From the overall light curves, the case with $\phi_0=\pi/2$ exhibits better distinguishability. This is because when $\phi_0=0$, after reaching the first peak intensity at $t=4.44$ min, the intensity decreases only for a short period before rapidly increasing to the second flare, and the decreased intensity is not very pronounced. Compared to the case with $\phi_0=\pi/2$, this makes it less likely to be clearly noticed by observers. Therefore, for subsequent observations, we adopt $\phi_0=\pi/2$, as it allows for easier distinction in the overall light curves. This seems to indicate that changing the observer's azimuthal angle alters the time interval between the first and second flares.

\section{Parameter Analysis}
In this section, we analyze the influence of the expansion parameter $r_0$ and the spin $a$ on the hot spot images. Compared to the Kerr black hole, the Kerr-Sen black hole includes an expansion parameter $r_0$ related to charge. Therefore, we first analyze the impact of the expansion parameter on the hot spot intensity. We set the parameters to $r_0=0.05, 0.08, 0.1$, corresponding to $\varepsilon_{-}=-0.4178, -0.4333, -0.4254$ and $\varepsilon_{+}=11.4789, 10.3194, 9.8258$, respectively. All cases satisfy the energy extraction condition. All other parameters remain the same as in Section 4.A.
\begin{figure}[!h]
  \centering
  \begin{subfigure}{0.3\textwidth}
    \centering
    \includegraphics[width=\linewidth]{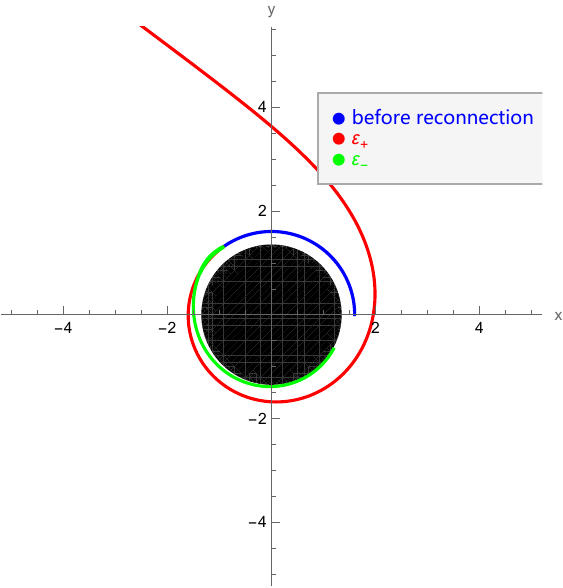}
    \caption{$r_0=0.05$}
    \label{fig:5a}
  \end{subfigure}
  \begin{subfigure}{0.3\textwidth}
    \centering
    \includegraphics[width=\linewidth]{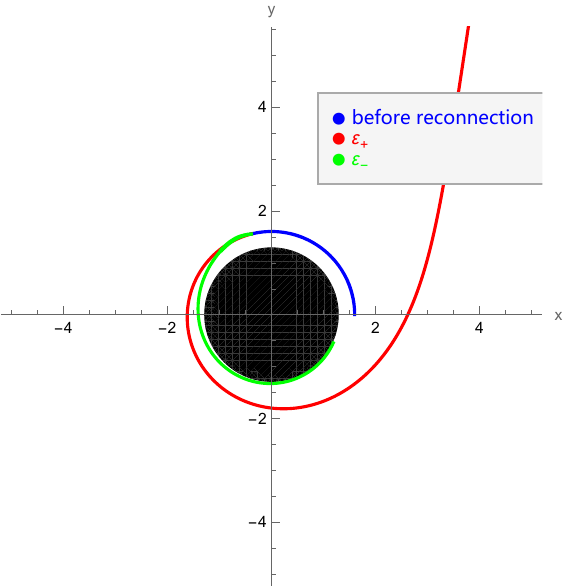} 
    \caption{$r_0=0.08$}
    \label{fig:5b}  
  \end{subfigure}
  \begin{subfigure}{0.3\textwidth}
    \centering
    \includegraphics[width=\linewidth]{2a.pdf} 
    \caption{$r_0=0.1$}
    \label{fig:5c}  
  \end{subfigure}
  \begin{subfigure}{0.3\textwidth}
    \centering
    \includegraphics[width=\linewidth]{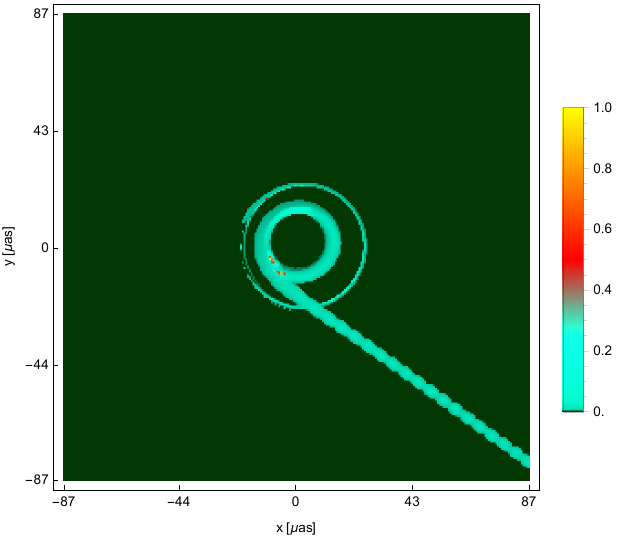} 
    \caption{$r_0=0.05$}
    \label{fig:5d}  
  \end{subfigure}
  \begin{subfigure}{0.3\textwidth}
    \centering
    \includegraphics[width=\linewidth]{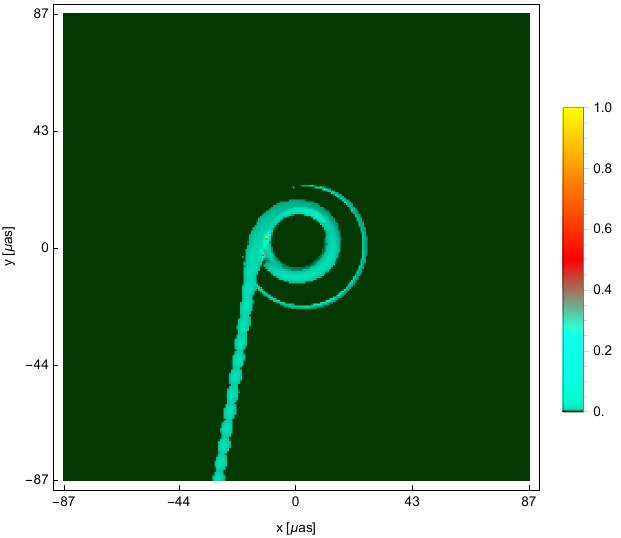} 
    \caption{$r_0=0.08$}
    \label{fig:5e}  
  \end{subfigure}
   \begin{subfigure}{0.3\textwidth}
    \centering
    \includegraphics[width=\linewidth]{2b.pdf} 
    \caption{$r_0=0.1$}
    \label{fig:5f}  
  \end{subfigure} 
    \begin{subfigure}{0.3\textwidth}
    \centering
    \includegraphics[width=\linewidth]{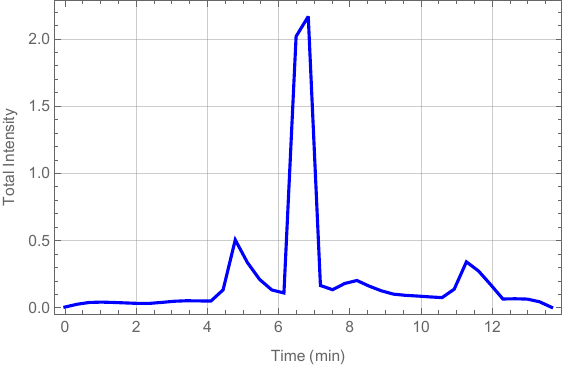} 
    \caption{$r_0=0.05$}
    \label{fig:5g}  
  \end{subfigure} 
   \begin{subfigure}{0.3\textwidth}
    \centering
    \includegraphics[width=\linewidth]{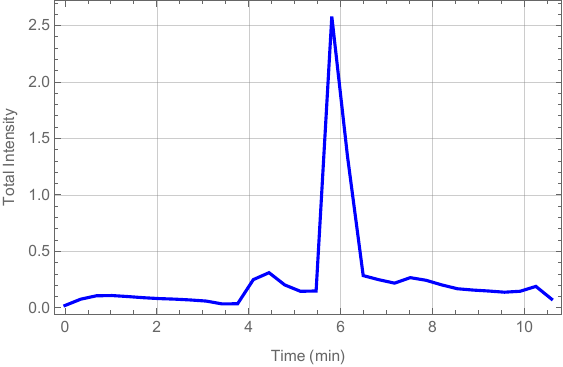} 
    \caption{$r_0=0.08$}
    \label{fig:5h}  
  \end{subfigure} 
   \begin{subfigure}{0.3\textwidth}
    \centering
    \includegraphics[width=\linewidth]{2c.pdf} 
    \caption{$r_0=0.1$}
    \label{fig:5i}  
  \end{subfigure} 
  \begin{subfigure}{0.3\textwidth}
    \centering
    \includegraphics[width=\linewidth]{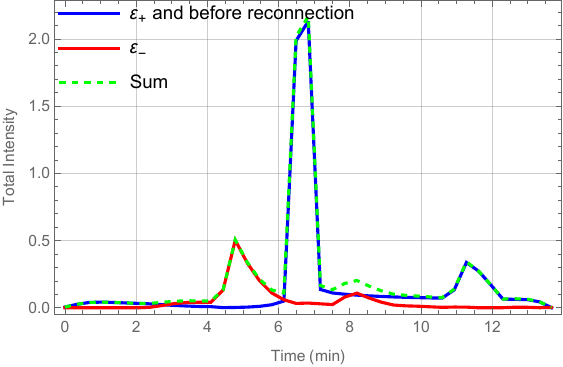} 
    \caption{$r_0=0.05$}
    \label{fig:5j}  
  \end{subfigure} 
  \begin{subfigure}{0.3\textwidth}
    \centering
    \includegraphics[width=\linewidth]{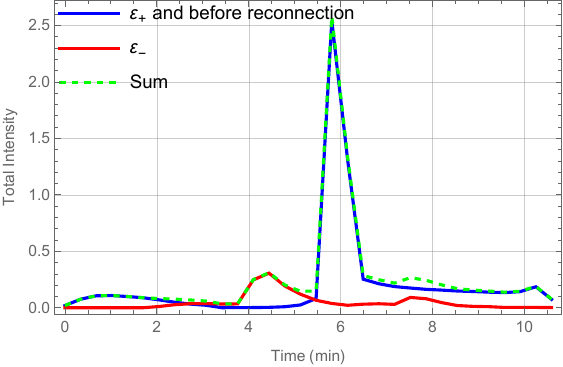} 
    \caption{$r_0=0.08$}
    \label{fig:5k}  
  \end{subfigure} 
  \begin{subfigure}{0.3\textwidth}
    \centering
    \includegraphics[width=\linewidth]{3b.pdf} 
    \caption{$r_0=0.1$}
    \label{fig:5l}  
  \end{subfigure} 
  \caption{For different values of $r_0$, the first row shows the trajectories of the plasmoid, the second row depicts the normalized intensity distribution of the hot spot observed, the third row displays the light curves of the hot spot emission, and the fourth row presents the light curves for the three cases.}
  \label{fig:5}
\end{figure}
From Fig. \ref{fig:5}, it can be observed that for different values of $r_0$, the first flare is consistently produced by the decelerated plasmoid. This further indicates that the first flare has the potential to serve as a signature of ongoing energy extraction via the Penrose process. It can also be seen that as $r_0$ increases, the total intensity of the first flare gradually decreases. This suggests that a larger expansion parameter makes it more difficult to identify the energy extraction signal.

Next, we analyze the influence of spin on the hot spot intensity. We set $r_0=0.05$. According to equation \eqref{3}, the extremal black hole in this case corresponds to $a=0.975$. We select spin parameters $a=0.9, 0.93, 0.97$, corresponding to $\varepsilon_{-}=-0.4178, -0.5326, -0.6198$ and $\varepsilon_{+}=11.4789, 9.8280, 8.8217$, respectively. All cases satisfy the energy extraction condition. Apart from $r_0$ and $a$, all other parameters remain the same as in Section 4.A.
\begin{figure}[!h]
  \centering
  \begin{subfigure}{0.3\textwidth}
    \centering
    \includegraphics[width=\linewidth]{5a.pdf}
    \caption{$a=0.9$}
    \label{fig:6a}
  \end{subfigure}
  \begin{subfigure}{0.3\textwidth}
    \centering
    \includegraphics[width=\linewidth]{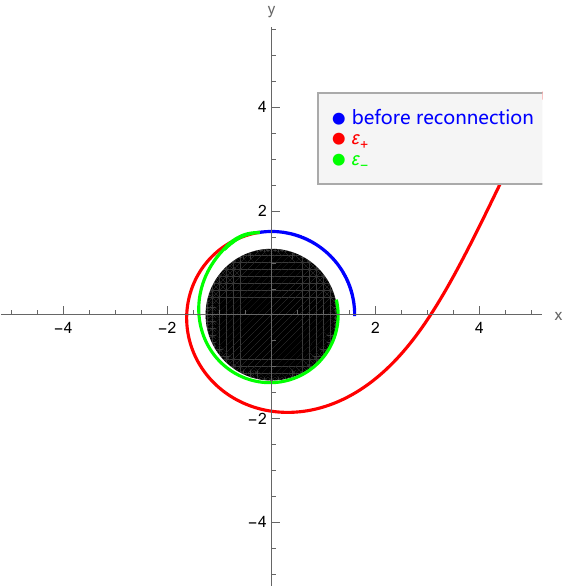} 
    \caption{$a=0.93$}
    \label{fig:6b}  
  \end{subfigure}
  \begin{subfigure}{0.3\textwidth}
    \centering
    \includegraphics[width=\linewidth]{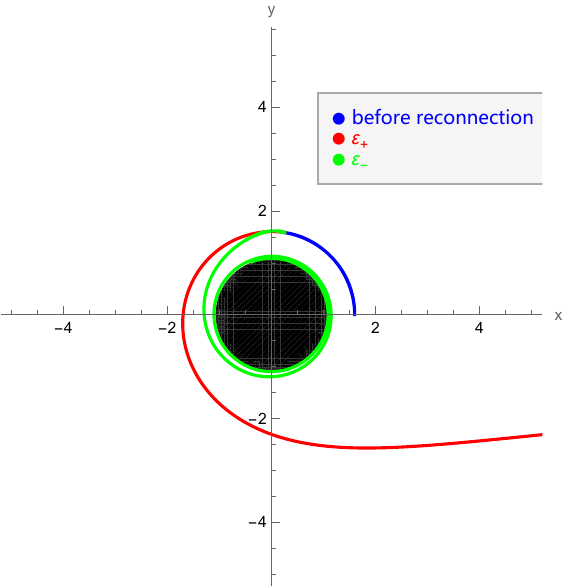} 
    \caption{$a=0.97$}
    \label{fig:6c}  
  \end{subfigure}
    \begin{subfigure}{0.3\textwidth}
    \centering
    \includegraphics[width=\linewidth]{5d.pdf} 
    \caption{$a=0.9$}
    \label{fig:6d}  
  \end{subfigure}
  \begin{subfigure}{0.3\textwidth}
    \centering
    \includegraphics[width=\linewidth]{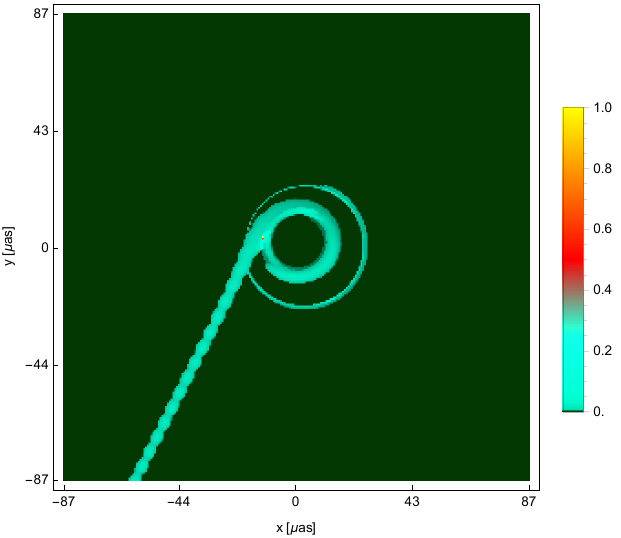} 
    \caption{$a=0.93$}
    \label{fig:6e}  
  \end{subfigure}
   \begin{subfigure}{0.3\textwidth}
    \centering
    \includegraphics[width=\linewidth]{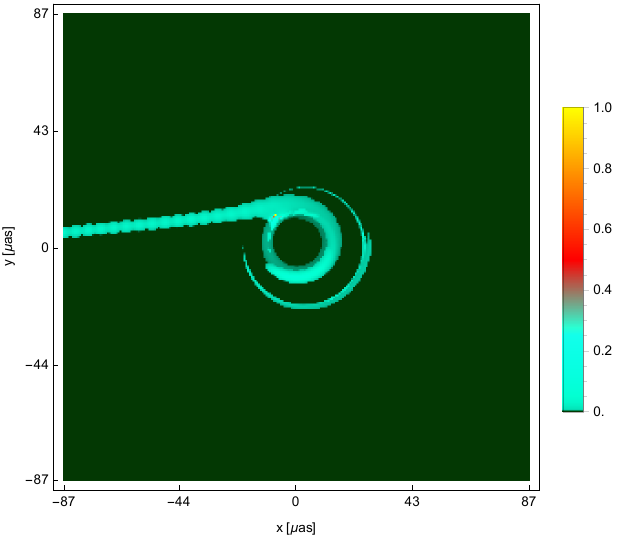} 
    \caption{$a=0.97$}
    \label{fig:6f}  
  \end{subfigure} 
    \begin{subfigure}{0.3\textwidth}
    \centering
    \includegraphics[width=\linewidth]{5g.pdf} 
    \caption{$a=0.9$}
    \label{fig:6g}  
  \end{subfigure} 
   \begin{subfigure}{0.3\textwidth}
    \centering
    \includegraphics[width=\linewidth]{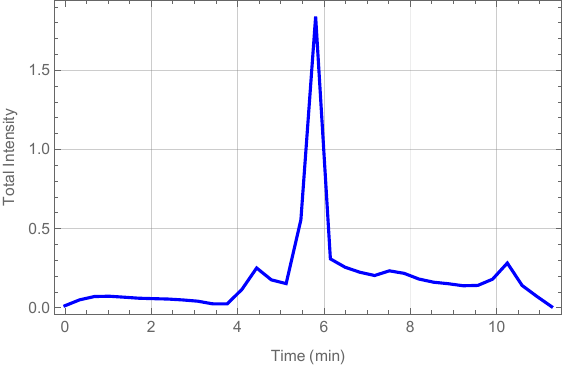} 
    \caption{$a=0.93$}
    \label{fig:6h}  
  \end{subfigure} 
   \begin{subfigure}{0.3\textwidth}
    \centering
    \includegraphics[width=\linewidth]{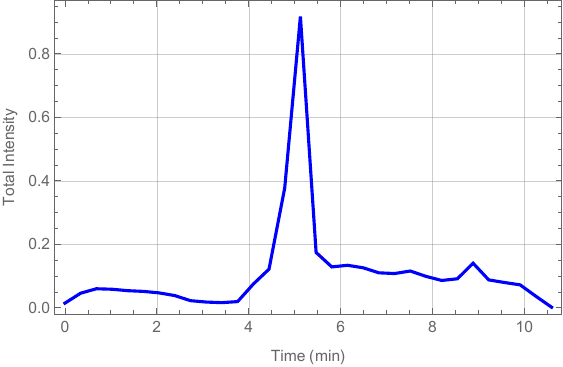} 
    \caption{$a=0.97$}
    \label{fig:6i}  
  \end{subfigure} 
  \begin{subfigure}{0.3\textwidth}
    \centering
    \includegraphics[width=\linewidth]{5j.pdf} 
    \caption{$a=0.9$}
    \label{fig:6j}  
  \end{subfigure} 
  \begin{subfigure}{0.3\textwidth}
    \centering
    \includegraphics[width=\linewidth]{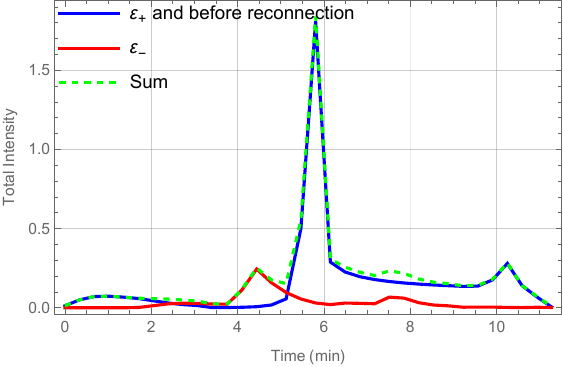} 
    \caption{$a=0.93$}
    \label{fig:6k}  
  \end{subfigure} 
  \begin{subfigure}{0.3\textwidth}
    \centering
    \includegraphics[width=\linewidth]{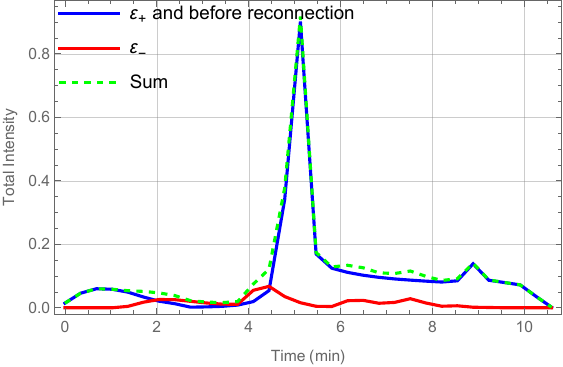} 
    \caption{$a=0.97$}
    \label{fig:6l}  
  \end{subfigure} 
\caption{For different values of $a$, the first row shows the trajectories of the plasmoid, the second row depicts the normalized intensity distribution of the hot spot observed, the third row displays the light curves of the hot spot emission, and the fourth row presents the light curves for the three cases.}
  \label{fig:6}
\end{figure}

From Fig. \ref{fig:6}, it can be observed that as the spin increases, the first flare produced by $\varepsilon_{-}$ gradually weakens, and the time interval between the first and second flares becomes progressively shorter. This results in the total light curve for the near-extremal case of $a=0.97$ showing only two discernible flares. Based on the analysis in Section 4.C, adjusting the observer's azimuthal angle might help increase the time interval between the two flares. However, the key issue is that in the near-extremal case of $a=0.97$, the intensity bump produced by $\varepsilon_{-}$ is extremely faint. Even if the time interval between flares is increased, this feature may not be perceptible as a distinct flare. The above analysis indicates that as the spin increases, identifying the energy extraction signal becomes more challenging.

\section{Conclusion}

This paper employs the hot spot imaging method to observe the motion of plasma after (and briefly before) the occurrence of the Comisso-Asenjo process in a Kerr-Sen black hole. First, we provided a brief review of the Comisso-Asenjo process. Then, we introduced our hot spot model and imaging methodology. Subsequently, we presented the observational results for cases with energy extraction, without energy extraction, and with varying observer azimuthal angles. Finally, we discussed the influence of the expansion parameter and spin on the hot spot imaging. Our research demonstrates that during the hot spot imaging process, observers detect three flares within the observation period. The first flare has the potential to serve as a signature of ongoing energy extraction. Changing the observer's azimuthal angle may alter the time interval between the first and second flares. A larger charge (associated with a larger expansion parameter $r_0$) leads to a lower total intensity of the first flare, thereby making it more difficult to identify the energy extraction signal. A higher spin also results in a lower total intensity of the first flare and a shorter time interval between the first and second flares, further complicating the identification of the energy extraction signal. Notably, for near-extremal black holes, only two flares are typically observable, preventing the clear identification of the energy extraction signal. However, numerous current studies indicate that achieving high power and high efficiency in energy extraction requires the spin to be as close as possible to the extremal value. This suggests that reconciling the requirements for high power/efficiency energy extraction with the clear identification of its observational signature necessitates further research within other models or frameworks. We anticipate that future work might enable the identification of the energy extraction signal even for near-extremal black holes under similar parameters. It is important to note that throughout this paper, we fixed many parameters such as $\sigma$, $\xi$, $s$, $\theta_0$, $r_X$, and the observer's radial distance. If the constraints on these parameters were relaxed, it might be possible to observe three flares even for near-extremal cases. However, under the specific parameters used in this study for near-extremal black holes undergoing energy extraction, only two flares are observable. This indicates that while the occurrence of energy extraction might be signaled by a faint first flare preceding the brighter ones, this first flare may not always be detectable. We intend to explore this aspect further in future discussions.

Our hot spot model has not yet incorporated the emissivity of the hot spot across different frequency bands, which is an aspect requiring further investigation. Furthermore, considering the complex astrophysical environment around black holes, the trajectories of hot spots might not strictly follow geodesic motion. This provides a clear direction for our future research. The study of the Penrose process is advancing rapidly \cite{37,38,39}. Future research could also consider employing different mechanisms that trigger the Penrose process for hot spot imaging, aiming to identify more and clearer signatures of energy extraction.

\noindent {\bf Acknowledgments}

\noindent
This work is supported by the National Natural Science Foundation of China (Grants Nos. 12375043,
12575069 ). We thank Zhixing Zhao, Fan Zhou, and Minyong Guo for useful discussions.

\end{document}